\documentclass{article} 

\newtheorem{proposition}{Proposition}
\newtheorem{definition}{Definition}

\begin{document}

\title{Closed Universes With Black Holes But No
Event Horizons As a Solution to the Black
Hole Information Problem\thanks{This work was
supported in part by the National Science Foundation
under grant number DMS-97-32058.}}

\author{Frank J. Tipler\thanks{e-mail
address: FRANK.TIPLER@TULANE.EDU}\, and Jessica
Graber\\ Department of Mathematics and Department
of Physics \\ Tulane University \\ New Orleans, LA
70118 USA \\
\\Matthew McGinley\thanks{Department of
Mathematics, University of Wisconsin, Madison, WI
53706 USA}, Joshua Nichols-Barrer\thanks{Department
of Mathematics, Harvard University, Cambridge MA
02138 USA}, \\and Christopher Staecker\thanks{
Department of Mathematics, Bates College, Lewiston, ME
04240}\\ Department of Mathematics\\Tulane
University\\New Orleans, LA 70118 USA}

\maketitle

\begin{abstract}
We show it is possible for the
information paradox in black hole evaporation to be
resolved classically.  Using standard junction
conditions, we attach the general closed spherically
symmetric dust metric to a spacetime satisfying all
standard energy conditions but with a single point future
c-boundary.  The resulting Omega Point spacetime,
which has NO event horizons, nevertheless has black
hole type trapped surfaces and hence black holes. 
But since there are no event horizons, information
eventually escapes from the black holes.  We show that
a scalar quintessence field with an appropriate
exponential potential near the final singularity would
give  rise to an Omega Point final singularity.
\end{abstract}

PACS numbers: 04.60.+n, 0470.Dy, 97.60.Lf
\pagebreak

\section{Introduction}
\setcounter{equation}{0}

One of the outstanding questions of black hole physics
is determining what happens to the information that
falls into a black hole.  Hawking has shown
\cite{Hawking1} that a black hole radiates away its
mass, and he pointed out that if a black hole were to
completely evaporate --- which it inevitably will in a
universe that exists forever in cosmological proper time 
(a Planck size remnant would probably be inconsistent
with the Bekenstein Bound (\cite{Beken1}, \cite{Beken2},
\cite{wald2}).) --- then any information exclusively
inside the black hole would disappear from the
universe, violating unitarity.  Many solutions have been
proposed to resolve this paradox.  Hawking himself
believes that unitarity is indeed violated, but it has
been argued that such a resolution would be
inconsistent with locality and/or conservation of energy
\cite{Susskind3}.  Susskind (\cite{Susskind1},
\cite{Susskind2}) and 't Hooft \cite{tHooft1} propose
that all information inside a black hole is also
completely encoded on its surface, so there is no net
information inside the black hole.  But at the
semi-classical level, this ``holographic principle'' would
not resolve the paradox, because the generators of an
event horizon --- the black hole surface --- cannot end
in spacetime, but at a singularity which itself would
annihilate any information on the horizon.  To avoid
unitarity violation, the information must get outside the
black hole event horizon.

Bekenstein \cite{Beken3} has pointed out that the
Hawking radiation is not quite a blackbody spectrum,
and thus it carries some information about the initial
state of the black hole.  He has shown that the permitted
information outflow rate can be as large as the rate of
black hole's entropy decrease, and hence it is possible
for information to gradually leak out of a black hole
during evaporation.  However, Bekenstein emphasized
that he had not demonstrated that {\it all} the
information got out, just that it was possible that it
did, and if only one bit of information fails to escape,
unitarity will be violated.  Bekenstein also did not
address the semi-classical event horizon problem.

We shall show in this paper that a purely classical
gravity solution to the black hole information problem
is possible and consistent with all observations:  the
universe may have no event horizons at all.  In such a
universe, there would be no black hole event horizons to
prevent the exchange of information between one part
of the cosmos and another.  A spacetime with no event
horizons has a future c-boundary (\cite{hawk}, pp.
217--221) which is topologically a single point, and
hence has been called \cite{tipler3} an {\it Omega
Point Spacetime}.  It can be shown \cite{tipler3} that if
a spacetime's future c-boundary is a single point, then
the spacetime necessarily admits compact Cauchy
surfaces, and the global spacetime topology is $S\times
R^1$, where $S$ is the topology of any Cauchy surface. 
Even in a universe with compact Cauchy surfaces, we
would expect black holes to evaporate to completion if
the universe were to expand forever.  Hence a
spacetime which avoids the black hole information
paradox because of the absence of event horizons
would have to end in a final singularity before any
black hole would have time to evaporate.  Since the
expected black hole lifetime is
$10^{64}(M/M_\odot)^3$ years, \cite{wald2}, our
universe would have to expand to a maximum size and
recontract before $10^{64}$ years have passed.  It can
be shown (\cite{tipler}, \cite{tipler4}) that the only
two simple topologies possible for universe with a
maximal Cauchy hypersurface and satisfying the weak
energy condition are $S^3$ and
$S^2\times S^1$.

We shall construct in this paper a spherically symmetric
 $S^3$ universe with a black hole but with no event
horizons.  The spacetime will be shown to satisfy all
the standard energy conditions.  Indeed, the stress
energy tensor for the spacetime with be just
pressure-free dust in its expanding phase.  We shall
discuss various definitions of ``black hole'' in closed
universes, and show that the spacetime we construct
has a black hole by any of these definitions.  The
parameters of the constructed spacetime can be
chosen so that the black hole is identical to a black
hole in any dust spherically symmetric (Tolman-Bondi)
$S^3$ universe --- and hence it would be in appearance
a black hole with event horizons according to any
observations that could be carried out in the expanding
phase of a closed universe.  The null generators of
what is apparently the event horizons stay close to
the trapped surfaces during the expanding phase of the
closed universe, and only expand out into the universe
at large very close to the final singularity.  This means
the standard astrophysical analysis of black holes and
their collisions (e.g., \cite{matzner1}, \cite{matzner2},
\cite{price1}) can be trusted, since they are in the short
run the same as in asymptotically flat spacetime.  Thus
our proposal is quite different from many proposals
which eliminate event horizons by eliminating black
hole type trapped surfaces. In our proposal, black hole
trapped surfaces exist as usual, but they do not give rise
to event horizons.

Our paper will be organized as follows.  In Section 2
we shall construct a Friedmann-Robertson-Walker
(FRW) universe which is a standard dust FRW
closed universe until arbitrarily near the final
singularity when we join it to a metric which satisfies
all the energy conditions, which has a final
singularity, but which has no event horizons.  In
Section 3 we show that the no event horizon metric
constructed in Section 2 satisfies the Einstein
equations for a scalar field with a suitably chosen
exponential potential.  In Section 4, we shall
generalize the FRW $w=-1/3$ universe to the
spherically symmetric case, obtaining an
inhomogeneous (but spherically symmetric) spacetime
which satisfies all the standard energy conditions, yet
has a future c-boundary which is a single point.  In
Section 5, we join this modified version of the FRW event
horizonless metric to a general Tolman-Bondi closed
universe.  In Section 6, we discuss various definitions
for a black hole in a closed universe, and show that the
Tolman-Bondi universe parameters of the metric in
section 5 can be chosen so that by any of these
definitions, the expanding phase of the universe has a
black hole.  In Section 7, we show that the recent
supernova observations which strongly suggest that
the universe is currently accelerating are consistent
with a universe which recollapses to a final singularity
before any black hole has time to completely evaporate,
provided the acceleration is due to quintessence with
certain specified properties.  Finally in our concluding
Section 8, we shall point out how our ``no event
horizon'' solution to the black hole information paradox
naturally complements the ``holographic principle''
resolution, which assumes that all information in a
black hole interior is coded also on its surface.

\section{A 3-sphere FRW Universe with Final
Singularity But No Event Horizons}\label{Omegafrw}
\setcounter{equation}{0}

The Friedmann equation for an $S^3$ closed universe is

\begin{equation}
\left({1\over R}{dR\over dt}\right)^2 =
{8\pi GM\over3R^{3(1+w)}} -{1\over
R^2}\label{friedeq} \end{equation}

\noindent
where the pressure $p=w\rho$, with $w = \gamma-1$,
where $\gamma$ is the adiabatic index and $\rho$ the
mass density.  If $w =-1/3$, then 

\begin{equation}
R(t) = \sqrt{(8\pi GM/3)-1}(t_f-t)\label{scale}
\end{equation}

\noindent
is a solution to (\ref{friedeq}) for $t<t_f$ with a final
singularity at $t=t_f$, provided $(8\pi GM/3) >1$.  The
second order equation for the Friedmann universe,

\begin{equation}
{1\over R}{d^2R\over dt^2} = -{4\pi
G\over3}\left(\rho +3p\right)\label{2ndfriedeq}
\end{equation}

\noindent
is automatically satisfied for $p=-(1/3)\rho$ and
$d^2R/dt^2 = 0$.
 
The closed FRW universe with the scale factor
(\ref{scale}), namely \begin{equation}
\label{frw}
ds^2=-dt^2+R_0^2(t_f-t)^2[d\chi^2+\sin^2\chi(d\theta^2
+\sin^2\theta d\varphi^2)], \end{equation}
 
\noindent
has no event horizons; that is, its future c-boundary
consists of a single point --- the Omega Point.  Indeed,
the equation for future directed null geodesics, $ds^2=0$
can be integrated for radial null geodesics to give

\begin{equation}
\Delta\chi = \int^{t_f}{dt\over R(t)} =
+\infty\label{circum} \end{equation}

\noindent
which shows that radial null geodesics circumnavigate
the universe an infinite number of times as the future
c-boundary at $t=t_f$ is approached.  By homogeneity
and isotropy, we can transpose the coordinate system
so that any spatial location $(\chi,\theta,\phi)$ can
reach any other location $(\chi',\theta',\phi')$ via a {\it
radial} null geodesic segment, and (\ref{circum}) shows
such radial geodesics can be exchanged an infinite
number of times.  Hence all future endless timelike
curves define the same c-boundary point: the future
c-boundary is a single point.

A perfect fluid with $w=-1/3$ satisfies the weak,
the strong, and the dominant energy conditions, since
Hawking and Ellis have shown (\cite{hawk}, pp. 89-95)
that for diagonalizable stress energy tensors (Type I
matter), the weak energy condition will hold if $\rho
\geq 0$ and $\rho + p \geq 0$; the strong energy
condition will hold if $\rho + p \geq 0$ and $\rho + 3p
\geq 0$; and the dominant energy condition will hold if
$\rho \geq 0$ and $-\rho \leq p \leq \rho$.

 It is possible to join the metric with scale factor
(\ref{scale}) to any closed FRW universe at {\em any}
time in the collapsing phase.  Consider for example the
dust ($w=0$) scale factor:

\begin{eqnarray}
R(\tau) = {R_{max}\over2}(1-\cos\tau)\label{dustR}\\
t(\tau) = {R_{max}\over2}(\tau - \sin\tau)\label{dustt}
\end{eqnarray}

\noindent
where $0<\tau<2\pi$ is the conformal time, and
$R_{max}$ is the radius of the universe at maximum
expansion, which occurs when $\tau =\pi$.  We will
make the join at conformal time $\pi\leq t_{join}<
2\pi$, which by (\ref{dustt}) gives a proper time of

$$t_{join} \equiv t_{join}(\tau_{join}) =
{R_{max}\over2}(\tau_{join} - \sin\tau_{join})$$

The standard junction conditions require continuity of
the metric and its first derivatives at the join:

\begin{eqnarray}
R(\tau_{join}) = {R_{max}\over2}(1-\cos\tau_{join})
= R(t_{join}) = -Qt_{join} + A\\ dR/dt|_{\tau_{join}} =
\left({{dR/d\tau}\over{dt/d\tau}}\right)\Big|_{
\tau_{join}} ={\sin\tau_{join}\over{1-\cos\tau_{join}}} =
{dR/ dt}|_{t_{join}} = -Q
\end{eqnarray}

\noindent
where $Q\equiv \sqrt{8\pi GM/3 - 1}$ and $A\equiv
t_fQ$.  Solving for $t_f$ and $M$ yield

\begin{eqnarray}
{8\pi GM\over3} = {2\over{1-\cos\tau_{join}}}\\
t_f = {R_{max}\over2}\left(\tau_{join} +
{2[\cos\tau_{join} - 1]\over\sin\tau_{join}}\right)
\end{eqnarray}

Notice that as $\tau_{join} \rightarrow \pi^+$, we have
$t_f \rightarrow +\infty$, which means joining at the
time of maximum expansion would yield the Einstein
static universe thereafter.  As $\tau_{join} \rightarrow
2\pi$, we have $t_f \rightarrow \pi R_{max}$ and
$M\rightarrow +\infty$, which means that an
arbitrarily large total mass of the $w=-1/3$ matter
is required to eliminate event horizons if the join is
made arbitrarily close to the usual proper time end of a
dust FRW universe, $t_f = \pi R_{max}$.

The standard junction conditions yield a global metric
which is $C^\infty$, except at the join, where it is
$C^1$.  One can smooth this metric to one which is
$C^\infty$ everywhere and which satisfies the energy
conditions everywhere by allowing $w$ to vary
smoothly from $0$ to $-1/3$ in a neighborhood
$[t_{join}, t_{join}-\Delta t)$.  By the constraint FRW
equation
$(R^{-1}dR/dt)^2 = - R^{-2} + 8\pi G\rho/3$ and the
dynamical FRW equation $2R^{-1}d^2R/dt^2 = -
(R^{-1}dR/dt)^2 - R^{-2} -8\pi Gp$, continuity in $p$ and
$\rho$ would insure that $R(t)$ is $C^2$, and repeatedly
differentiating the dynamical equation would yield that
$R(t)$ is $C^\infty$ (the constraint and dynamical FRW
equations imply the conservation equation $T^{\mu\nu}
\,_{;\nu} =0$).

Since $ R^{\mu}\, _{\mu} = -8\pi GT^{\mu}\, _{\mu}
=16\pi G\rho = 16\pi GM/R^2$ for the $w=-1/3$
equation of state, where $ R^{\mu}\, _{\mu}$ is the
Ricci scalar, the Omega Point singularity at $t=t_f$ (at
$R=0$) is a p.p. curvature singularity (\cite{hawk}, p.
260).  The spacetime is a counter-example to a
conjecture by R.K. Sachs \cite{sachs} that the only
Omega Point spacetimes are formed by suitably
identifying Minkowski space, which would have locally
extendible singularities.

\section{A $w= -1/3$ Perfect Fluid Can Be
Generated By a Quintessence Scalar Field
With an Exponential Potential}\label{exp}
\setcounter{equation}{0}

\bigskip
We shall now show that a scalar field with exponential
potential will generate, at least in a FRW universe, a
$w= -1/3$ perfect fluid behavior near the final
singularity.   That is, the $w= -1/3$ perfect
fluid behaviour will be seen if the potential for the
scalar field $\phi$ is of the form $V(\phi) =
V_0e^{B\phi}$, where $V_0$ and $B$ are constants. 
Such a potential is often discussed as a particularly
plausible potential for the inflaton field which is
thought to be responsible for inflation in the early
universe, and as a model of the quintessence field
which is responsible for the cosmological
acceleration in the present epoch.  This will show that
a $w= -1/3$ equation of state is physically plausible
near the final singularity of a closed universe, and thus
that the absence of event horizons is physically
possible.

The stress energy tensor for a scalar field $\phi$ with
potential $V(\phi)$ is \cite{Turner3}

\begin{equation}
T_{\alpha\beta} =
\left[\phi_{;\alpha}\phi_{;\beta} -
{1\over2}g_{\alpha\beta}\left(\phi_{;\mu}\phi_{;\nu}
g^{\mu\nu} + 2V(\phi)\right)\right]
\end{equation}

In the FRW universe, we have $\phi = \phi(t)$, so
$\phi_{;i} =0$ and $\phi_{;0} = \phi_{,0}$, where the $i$
denotes a spatial coordinate, $0$ the time coordinate
$t$, and the semicolon and comma denote the covariant
and partial derivatives respectively.  In a local
orthonormal frame we obtain

\begin{equation}
T_{{\hat0}{\hat0}} ={1\over2}(\phi_{,\hat0})^2
+ V(\phi)
\end{equation}

\noindent
and

\begin{equation}
T_{{\hat0}{\hat0}} + 3T_{{\hat i}{\hat i}} =
2\left[(\phi_{,\hat0})^2 - V(\phi)\right]
\end{equation}

If $w = -1/3$, $T_{{\hat0}{\hat0}} + 3T_{{\hat i}
{\hat i}} =0$, which means

\begin{equation}
V(\phi) = (\phi_{,\hat0})^2
=(\phi_{,0})^2
\end{equation}

\noindent
where we have used $\phi_{,\hat0} = \phi_{,0} =
d\phi/dt$.  Thus

\begin{equation}
8\pi GT_{{\hat0}{\hat0}} = 12\pi G(\phi_{,0})^2 =
G_{{\hat0}{\hat0}} = {{3((R_{,0})^2 + 1)}\over R^2} =
{{3(R_0^2 +1)}\over {R_0^2(t_f - t)^2}}
\end{equation}

Taking the square root gives

\begin{equation}
{{d\phi}\over {dt}} = {{\sqrt{(R_0^2 +1)/4\pi G}}\over
{R_0(t_f - t)}}\label{phiprime}
\end{equation}

\noindent
which can be immediately integrated to yield

\begin{equation}
\phi_0 - \phi = \sqrt{(1/4\pi G)(1+ 1/R_0^2)}\ln(t_f
-t)\label{integralphi}
\end{equation}

\noindent
where $\phi_0$ is a constant.  Equation
(\ref{integralphi}) can be written

\begin{equation}
(t_f - t)^{-1} =
\exp\left[(\phi-\phi_0)/\sqrt{(1/4\pi G)
(1+1/R_0^2)}\right]\label{time}
\end{equation}

We thus obtain for the potential

\begin{equation}
V(\phi) = (\phi_{,0})^2 = {(R_0^2 +1)\over4\pi G
R_0^2}\left[{1\over(t_f-t)^2}\right] = V_0e^{B\phi}
\label{V(phi)}
\end{equation}

\noindent
where

\begin{equation}
B = \sqrt{{16\pi GR_0^2}\over{R_0^2
+1}}\label{constantB}
\end{equation}

\noindent
and

\begin{equation}
V_0 = {(R_0^2 +1)\over 4\pi GR_0^2}e^{-B\phi_0}
\label{constantVzero}
\end{equation}

It was pointed out in the previous Section that the
join between the dust (or radiation) dominated FRW
part of the universe and the $w= -1/3$ portion
can be made at any time and at any radius.  If the
constants $V_0$ and $B$ are fixed by the laws of
physics, then as the above relation between these
constants and the constants $R_0$ and $\phi_0$
indicate, the physical laws would also restrict the
radius of the join, and the value of the scalar field at
the join.

\medskip
It is interesting to confirm that the potential
(\ref{V(phi)}) satisfies the second order equation of
motion for a scalar field in the FRW universe with $R(t)
= R_0(t_f-t)$.  The equation of motion with arbitrary
scalar potential $V(\phi)$ is (\cite{tipler3}, p. 466;
\cite{tipler4}, p. 431):

\begin{equation}
\phi_{;\alpha}\,^{;\alpha} = {\partial
V(\phi)\over \partial\phi}\label{eqmotion1}
\end{equation}

In the FRW universe we have
$\phi_{;\alpha}\,^{;\alpha} = (\phi_{,0})\,^{;0} =
(\phi^{,0})\,_{;0} = (-\phi_{,0})\,_{;0}$, and in a
coordinate basis, the identity $A^\alpha\,_{;\alpha} =
(1/\sqrt{-g})(\sqrt{-g}A^\alpha)_{,\alpha}$, for any
vector field $A^\alpha$, applies.  Thus in a FRW
coordinate basis, the scalar field equation of motion
can be written

\begin{equation}
{1\over R^3}\left(R^3(-\phi_{,0})\right)_{,0} =
{\partial V(\phi)\over\partial\phi}
\label{eqmotion2}
\end{equation}

\noindent
which can be reduced to the standard expression
$\ddot\phi + 3H\dot\phi + V'(\phi) = 0$ as follows.  In
both a coordinate basis, and in an orthonormal basis,
we have $A_{,0} = dA/dt$, for any function $A$.  Thus,
for an exponential potential, we have $\partial
V(\phi)/\partial\phi = BV(\phi) = B(\phi_{,0})^2$. 
Using $R = R_0(t_f -t)$ and the expression
(\ref{phiprime}) for $\phi_{,0}$, it is confirmed that
(\ref{eqmotion2}) is indeed an identity.

\medskip
An alternative derivation of the fact that the
$w=-1/3$ equation of state can be generated by a
scalar field with exponential potential would be to
make use of Barrow's work (\cite{Barrow1},
\cite{Burd1}) on scalar fields with exponential potentials
in flat space (FRW k = 0).  Barrow in fact noted
\cite{Barrow1} that in the far future, an exponential
potential could give rise to the $w=-1/3$ equation of
state in the k = +1 case, but he did not attempt to derive
the constants ($B$ and $V_0$ above) that would allow
the $w=-1/3$ equation of state to be joined to a dust
equation of state for earlier times, which is why we did
the calculation above.  In addition, Vilenkin
\cite{Vilenkin} (see also \cite{Turner2}) has pointed out
that a $w=-1/3$ equation of state can be generated by a
tangled network of very light cosmic strings.

\medskip
In joining two metrics with different equations of
state, one effectively assumes that one form of
matter disappears and is replaced by the other.  More
realistically, if a scalar field were to be present near
a final singularity, we would expect it to be in addition
to dust or radiation already present.  In such a
situation, a pure exponential potential uncoupled to
the other forms of matter would {\it not} give rise to
a single c-boundary point, if its stress-energy
tensor increased as $R^{-2}$, since dust and radiation
would increase as $R^{-3}$ and $R^{-4}$ respectively;
such a universe would inevitably become radiation
dominated sufficiently near the singularity.  However,
in a FRW universe, we can always find, for any assumed
mixture of dust and radiation, a suitable potential
$V(\phi)$ which would have the effect of cancelling
out the gravitational force of the dust and matter
fields, leaving an effective pure exponential scalar
field (this is in effect what happens after the join
between the $w= -1/3$ equation of state and the $w =
0$ equation of state fluids).

\medskip
But the actual universe is not expected to be FRW near
the final singularity.  Even if the universe were
FRW in the beginning, we would expect it to become
curvature dominated near the final singularity, since
the ``effective energy density'' curvature perturbations
around FRW grow as $R^{-6}$, much faster than the
densities of dust or radiation (\cite{mtw}, p. 807). 
So in the actual universe, the elimination of event
horizons would have to be carried out by the global
collective interactions (of known forces) which give rise
to the Misner mixmaster horizon elimination
mechanism, as described in \cite{tipler3}.

\medskip
On the other hand, a pure scalar inflaton (quintessence
field) with exponential potential might be expected to
be the entire matter content in the very early universe,
and the initial singularity might be expected to be
FRW.  In such a case the effect of such an inflaton field
would be to eliminate the particle horizons.  In other
words, with an exponential inflaton (quintessence)
field, the horizon problem of cosmology would be
automatically resolved.

\section{Generalizing the FRW $w=-1/3$ Omega Point
Spacetime to the Spherically Symmetric
Case}\label{secNZ}
\setcounter{equation}{0}

The approach used in Section \ref{Omegafrw} for
creating spacetimes with no event horizons can be
generalized to yield a wider class of such spacetimes. 
Instead of using the metric (\ref{frw}), we introduce
functions $N(\chi)$ and $Z(\chi)$, where $N$ is positive
on $[0,\pi]$ and $Z$ is positive on $(0,\pi)$, vanishing at
$0$ and $\pi$.  The metric we then use is
\begin{equation}\label{NZ}
ds^2=-dt^2+(t_f-t)^2[N^2d\chi^2+
Z^2(d\theta^2+\sin^2\theta d\varphi^2)] \end{equation} 
{\proposition A Tolman-Bondi spacetime with metric
(\ref{NZ}) has a c-boundary which is a single point.}\\

\noindent {\em Proof}. To check that this spacetime
actually has no event horizons, we mimic the calculation
of the same proposition for the $w= -1/3$ 
FRW universe in section \ref{Omegafrw}.   
Let $N_{max}$ be the maximum value of
$N$ on $[0,\pi]$.  Then 
\begin{equation} \label {NZint}
\Delta\chi=\int^{t_f}\frac{dt}{N(\chi)(t_f-t)}\geq
\int^{t_f}\frac{dt}{N_{max}(t_f-t)}=+\infty.
\end{equation} 

Thus in this class of spacetimes, radial
null geodesics are capable of hitting every value of
$\chi$ an infinite number of times.  In order to conclude
that every point in space can communicate with every
other point, however, we must refine the argument
given in Section \ref{Omegafrw} a bit, for we no longer
have the symmetry of the 3-sphere to exploit.  We do,
however, still have (2-)spherical symmetry.  Therefore
we can say that a null geodesic may be sent from the
origin to any
$(\chi,\theta,\varphi)$, and vice versa.  Hence, given
points
$P_1=(\chi_1,\theta_1,\varphi_1)$ and
$P_2=(\chi_2,\theta_2,\varphi_2)$ which desire to
communicate with one another, there exists a piecewise
$C^\infty$ null curve from $P_1$ to $P_2$, consisting of
a null curve from $P_1$ to the origin and then a null
curve from the origin to $P_2$.  Applying an
elementary result of Penrose (\cite{penrose}, Lemma
2.16), we conclude that there exists a timelike or null
curve from $P_1$ to $P_2$, which is precisely what we
wanted.  QED.

We would like the spacetime (\ref{NZ}) to satisfy the
weak, dominant, and strong energy conditions
\cite{hawk}.  Let $G$ be the Einstein tensor of this
spacetime.  Using the equations for the nonzero
components of the Einstein tensor in \cite{exact}, we
can compute $G$ in the orthonormal basis
$\omega^{\hat \imath}$, where: $$\omega^{\hat
0}=dt,\quad\omega^{\hat
1}=N(t_f-t)d\chi,\quad\omega^{\hat
2}=Z(t_f-t)d\theta,\quad\omega^{\hat
3}=Z(t_f-t)\sin\theta d\varphi.$$ In this basis, all
off-diagonal terms of $G$ are zero.  Thus all matter is
Type I \cite{hawk}, and the energy conditions will hold
if the following six conditions are satisfied:
\begin{eqnarray} G^{\hat 0\hat 0}\geq 0\label{ene1}\\
G^{\hat 0\hat 0}+G^{\hat 1\hat 1}+G^{\hat 2\hat
2}+G^{\hat 3\hat 3}\label{ene2}\geq 0\\ G^{\hat 0\hat
0}+G^{\hat 1\hat 1}\geq 0\label{ene3}\\ G^{\hat 0\hat
0}-G^{\hat 1\hat 1}\geq 0\label{ene4}\\ G^{\hat 0\hat
0}+G^{\hat 2\hat 2}=G^{\hat 0\hat 0}+G^{\hat 3\hat
3}\geq 0\label{ene5}\\ G^{\hat 0\hat 0}-G^{\hat 2\hat
2}=G^{\hat 0\hat 0}-G^{\hat 3\hat 3}\geq 0\label{ene6}
\end{eqnarray}

The strong energy condition is (\ref{ene1}) and
(\ref{ene2}), the weak is (\ref{ene1}), (\ref{ene3}), and
(\ref{ene5}), and the dominant is (\ref{ene1}) and
(\ref{ene3})-(\ref{ene6}).  Computing these expressions
with our metric (\ref{NZ}), we see that
$G^{00}+G^{11}+G^{22}+G^{33}=0$ identically, and
(\ref{ene1}), (\ref{ene3}), (\ref{ene4}), (\ref{ene5}),
and (\ref{ene6}) are equivalent to, respectively:

\begin{eqnarray}
3+\frac{1}{Z^2}-\frac{1}{N^2}\Big(\frac{2Z''}{Z}
-\frac{2Z'N'}{ZN}+\frac{(Z')^2}{Z^2}\Big)\geq 0\\
2-\frac{2}{N^2}\Big(\frac{Z''}{Z}-\frac{Z'N'}{ZN}\Big)\geq
0\\
4+\frac{2}{Z^2}-\frac{2}{N^2}\Big(\frac{Z''}{Z}
-\frac{Z'N'}{ZN}+\frac{(Z')^2}{Z^2}\Big)\geq 0\\
2+\frac{1}{Z^2}-\frac{1}{N^2}\Big(\frac{Z''}{Z}
-\frac{Z'N'}{ZN}+\frac{(Z')^2}{Z^2}\Big)\geq 0\\
4+\frac{1}{Z^2}-\frac{1}{N^2}\Big(\frac{3Z''}{Z}-
\frac{3Z'N'}{ZN}+\frac{(Z')^2}{Z^2}\Big)\geq 0
\end{eqnarray} 

\noindent
where prime ($'$) denotes differentiation
with respect to $\chi$ (everything is a function of
$\chi$).  In the FRW case, $N=R_0$, $Z=R_0\sin\chi$,
and the energy condition equations are all are
equivalent to: $$1+\frac{1}{R_0^2}\geq 0.$$ In other
words, in the FRW case, the energy conditions are
always satisfied.

We shall now show that if the metric (\ref{NZ}) defines
a universe that is ``sufficiently large,'' it will
automatically satisfy the energy conditions.  Suppose we
are given functions $N_0(\chi)$ and $Z_0(\chi)$ such that
there exist constants $R_1,R_2,\epsilon_1,\epsilon_2>0$
with $N_0(\chi)=R_1$ and $Z_0(\chi)=R_1\sin\chi$ for
$0\leq\chi<\epsilon_1$ and $N_0(\chi)=R_2$ and
$Z_0(\chi)=R_2\sin\chi$ for $\pi-\epsilon_2<\chi\leq\pi$. 
In other words, $N_0$ and $Z_0$ look like the $N$ and $Z$
from FRW universes near $\chi=0$ and $\chi=\pi$. Then we
know that near $\chi=0$ and $\chi=\pi$, the energy
conditions are satisfied for $N=RN_0$ and $Z=RZ_0$, where
$R$ is an arbitrary positive constant.  Then since the
expressions multiplied by $\frac{1}{N^2}$ in the energy
conditions are bounded for $\chi\in
[\epsilon_1,\pi-\epsilon_2]$, we may find a constant
multiplier $R$ such that the metric (\ref{NZ}) with
$N=RN_0$ and $Z=RZ_0$ satisfies all the energy conditions
everywhere.  The current observational evidence
indicates that the universe is very close to being
spatially flat, so the actual universe satisfies the
``sufficiently large'' criterion.

\section{A 3-Sphere Universe Containing a Black
Hole But Having No Event Horizons}
\setcounter{equation}{0}

We will produce a large class of $S^3\times R^1$
spacetimes which are in their expanding phase, special
cases of the general spherically symmetric dust solution
\cite{exact} and which are eventually joined to a
spacetime of the type described in section \ref{secNZ},
so that they end with a c-boundary of a point (and
hence have no event horizons), and satisfy the energy
conditions everywhere.

\subsection{General Dust Solution}

The general spherically symmetric pressureless dust
solution \cite{exact} is:

\begin{equation}
ds^2=-dt^2+(1-f^2)^{-1}\Big(\frac{\partial
Y}{\partial\chi}\Big)_t^2d\chi^2+Y^2(d\theta^2
+\sin^2\theta d\varphi^2) 
\end{equation}

\noindent
where the notation $(\frac{\partial Y}{\partial\chi})_t$
denotes differentiation of $Y$ with respect to $\chi$
where the independent variables are $t$, $\chi$, $\theta$,
and $\varphi$ (subscripts to differentials in general will
specify independent variables, with the assumption that
$\theta$ and $\varphi$ are always independent), $f$ is an
arbitrary function of $\chi$ alone taking values in
$[0,1]$ and $Y$ and $t$ are given by:
\begin{eqnarray}
t=t_0(\chi)+(\eta-\sin\eta)\frac{m(\chi)}{f(\chi)^3}\label{tdef}\\
Y=(1-\cos\eta)\frac{m(\chi)}{f(\chi)^2}.
\end{eqnarray}
In the above expressions, $t_0$ is an arbitrary function of
$\chi$ alone, $m$ is another arbitrary function of $\chi$
positive on $(0,\pi)$, and $\eta$ is defined by
(\ref{tdef}).  The only restrictions on these free
functions are that to maintain the nondegeneracy of the
metric in a closed universe, $f$ should equal 1 at one
$\chi$-value in the interior of $[0,\pi]$, at which point
$m'$, $f'$ and $t'_0$ should all be zero. The general dust
metric becomes degenerate whenever $Y'=0$ and $f\neq 1$ or
$Y'\neq 0$ and $f=1$.  Such 2-spheres of degeneracy
correspond to shell-crossing singularities, and if these
degeneracy spheres occur before the final singularity at
$\eta = 2\pi$, they will give rise to a breakdown in
global hyperbolicity, as is well-known.  We shall assume
that the free functions $m,f,t_0$ are so chosen that this 
does not occur.

The dust case of the FRW metric (the case where
$w=0$) is a special case of this general metric.  Letting
$$f=\sin\chi, \qquad
m=\frac{R_{max}}{2}\sin^3\chi,\quad\mathrm{and}\quad
t_0=0,$$ one obtains
$$t=\frac{R_{max}}{2}(\eta-\sin\eta),\qquad
Y=\frac{R_{max}}{2}(1-\cos\eta)\sin\chi,$$
$$\mathrm{and}\quad\Big(\frac{\partial
Y}{\partial\chi}\Big)_t=\frac{R_{max}}{2}(1-\cos\eta).$$
The resulting metric is

$$ds^2=-dt^2+\Big[\frac{R_{max}}{2}(1-
\cos\eta)\Big]^2[d\chi^2+\sin^2\chi(d\theta^2+\sin^2\theta
d\varphi^2],$$ precisely the Friedmann collapsing dust
$S^3$ solution.

\subsection{The Join}

We have shown above that (\ref{frw}) can be joined in a
$C^1$ manner to any collapsing dust FRW $S^3$ universe at
{\em any} time in the collapsing phase by a suitable
choice of the constants $R_0$ and $t_f$.  We shall now
generalize this construction substantially, joining a
certain class of Tolman-Bondi pressureless dust solutions
(including the FRW $S^3$ collapsing dust solution) to
universes of the sort (\ref{NZ}), so that we produce a
large class of universes which start with pressureless
dust and in the end have no event horizons.

In making this join, we will allow the hypersurface
$\mathcal J$ along which the two metrics are joined to
vary as a free function.   For convenience, we will take
$\mathcal J$ to be spherically symmetric, parametrized as
$(t_{\mathcal J},\chi,\theta,\varphi)$ in the dust
universe, where $\eta_{\mathcal J}=\eta_{\mathcal
J}(\chi)$ is a free function of $\chi$, and 

\begin{equation}
t_{\mathcal J} \equiv t(\eta_{\mathcal
J}(\chi), \chi) = t_0(\chi) + (\eta_{\mathcal J} -
\sin\eta_{\mathcal J}) m(\chi)/f^3(\chi)
\label{eq:def.t.{J}}
\end{equation}

We will also assume that the $t$, $\chi$, $\theta$, and
$\varphi$ coordinates agree across $\mathcal J$. 
Therefore, we will have six degrees of freedom altogether:
$t_0$, $m$, $f$, $\eta_{\mathcal J}$, $N$, and $Z$, all
free functions of $\chi$.

To make the join $C^1$, we first must make it continuous
across $\mathcal J$.  This in particular means that the
metric coefficients will agree along $\mathcal J$ itself. 
Therefore the metric coefficients will agree in all
derivatives along vectors tangent to $\mathcal J$, and
thus in order to check that the join is $C^1$, one must
only check that the first derivatives of the metric
coefficients agree in a direction independent to the
tangent spaces of $\mathcal J$.  The direction we choose
is $(\partial/\partial t)_\chi$.  This direction is
linearly independent to the tangent spaces of $\mathcal J$
because $(\partial/\partial\eta)_\chi$ is never tangent to
$\mathcal J$ (since $\mathcal J$ is parametrized with
$\eta$ a function of $\chi$), and

$$\Big(\frac{\partial}{\partial
t}\Big)_\chi=\frac{f^3}{m(1 -
\cos\eta)}\Big(\frac{\partial}{\partial \eta}\Big)_\chi.$$

Since we are assuming that the coordinates are the same on
the dust universe as on the universe (\ref{NZ}), the
off-diagonal coefficients already agree (they are 0 on
both sides of $\mathcal J$), and $g_{tt}$ is -1 on both
sides of $\mathcal J$.  Furthermore, since we have
spherical symmetry on both sides of $\mathcal J$, we need
check only one of $g_{\theta\theta}$ and
$g_{\varphi\varphi}$.  We are left with {\em four}
junction conditions:

\begin{eqnarray}
\frac{(Y')^2}{1-f^2}=N^2(t_f-t_{\mathcal
J})^2\label{join1}&(g_{\chi\chi}\mathrm{\ continuous})\\
Y^2=Z^2(t_f-t_{\mathcal
J})^2\label{join2}&(g_{\theta\theta}\mathrm{\
continuous})\\ \frac{2Y'\dot
Y'}{1-f^2}=-2N^2(t_f-t_{\mathcal
J})\label{join3}&(g_{\chi\chi}\ C^1)\\ 2Y\dot
Y=-2Z^2(t_f-t_{\mathcal
J})\label{join4}&(g_{\theta\theta}\ C^1).
\end{eqnarray}

Here the dot ($\dot{}$) denotes application of
$(\partial/\partial t)_\chi$ and the prime ($'$) denotes
application of $(\partial/\partial \chi)_t$.  These
tangent vectors arise from the coordinate system
$(t,\chi,\theta,\varphi)$, and thus they commute with one
another, so that (\ref{join3}) makes sense.

Therefore we have six free functions --- $t_0(\chi),\,
m(\chi),\, f(\chi) $ on the Tolman-Bondi side of the join,
and $\eta_{\mathcal J} (\chi),\, N(\chi),\, Z(\chi)$ on
the final singularity side of the join --- and four
differential equations relating them.  One would expect
that these four equations would determine four of the
functions in terms of the other two, and this is exactly
what happens.  We find it convenient to choose $m(\chi)$
and $f(\chi)$ as the arbitrary functions, and expressing
all other functions in terms of these two.  After some
manipulation of the above equations we obtain

\begin{eqnarray}
\frac{(1+\cos\eta_{\mathcal J})^2}{|\sin^3\eta_{\mathcal
J}|}=C\frac{m}{f^3}\label{sol1}\\
t_0=t_f+\Big(\frac{2(1-\cos\eta_{\mathcal
J})}{\sin\eta_{\mathcal J}}-\eta_{\mathcal
J}\Big)\frac{m}{f^3}\label{sol2}\\ N=\frac{|\dot
Y'(\eta_{\mathcal J},\chi)|}{\sqrt{1-f^2}}\label{sol3}\\
Z=|\dot Y(\eta_{\mathcal J},\chi)|\label{sol4}.
\end{eqnarray}

\noindent
where $C$ is a constant of integration. 
Equation (\ref{sol1}) can be inverted to give
$\eta_{\mathcal J}(\chi)$.  Then $\eta_{\mathcal J}$ is
inserted into equations (\ref{sol2}), (\ref{sol3}), and
(\ref{sol4})to yield $t_0(\chi)$, $N(\chi)$, and $Z(\chi)$
respectively in terms of the arbitrary functions$m(\chi)$
and $f(\chi)$, and the constants $t_f$ and $C$ .  Notice
that the allowed Tolman-Bondi dust metrics are no longer
completely general, since the function $t_0$ is now fixed
rather than being completely arbitrary.  The constants
$t_f$ and $C$ allow the join to be made as far in the
future in proper time (the constant $t_f$) and in $\eta$
time (the constant $C$) as one wishes.  To see the latter,
note that equation (\ref{sol1}) is of the form
$F(\eta_{\mathcal J})=Cm/f^3$, where $F$ increases
monotonically from 0 to $+\infty$ as $\eta_{\mathcal J}$
ranges from $\pi$ to $2\pi$.  Thus if we make $C$
arbitrarily large, $\eta_{\mathcal J}$ can be made
arbitrarily close to $2\pi$, i.e. as close as we wish to
the final singularity.  Furthermore, one observes that the
boundary requirement of $t'_0$ (that it is 0 whenever $Y'$
and hence $m'$ and $f'$ are 0) is automatically satisfied,
since by the junction conditions

$$t'_0=\frac{m}{f^3}\Big(\frac{m'}{m}
-3\frac{f'}{f}\Big)\Big(\frac{(1-\cos\eta_{\mathcal
J})\sin\eta_{\mathcal J}}{2+\cos\eta_{\mathcal
J}}-\eta_{\mathcal J}+\sin\eta_{\mathcal J}\Big).$$

The join in Section \ref{Omegafrw} is in fact a special
case of this construction.  Consider the FRW choices for
$m$ and $f$:

$$m(\chi)=\frac{R_{max}}{2}\sin^3\chi,\quad\mathrm{and}
\quad f(\chi)=\sin\chi.$$ 
Then $m/f^3=1$ identically, so that by (\ref{sol1})
$\eta_{\mathcal J}$ will be a constant, and then by
(\ref{sol2}) $t_0$ will be constant.  Choosing $t_f$
appropriately, we may make $t_0$ identically zero, so that
our Tolman-Bondi universe is in fact the FRW collapsing
dust universe.  As noted above, we may make the join at
any time in the collapsing phase, just as in the FRW
construction of Section 2, and a simple calculation
reveals that the $N$ and $Z$ forced by the join are
precisely those which give the $w=-1/3$ universe.

\subsection{Join With a Possible ``Weak'' Shell-Crossing
Singularity}

In order to make the metric coefficients differentiable
across $\mathcal J$, we had to impose four conditions on
six functions.  We would, however, like to join a
completely arbitrary Tolman-Bondi metric to the metric
(\ref{NZ}), and this will require eliminating one of the
equations.  The junction condition which on physical
grounds is the least important is (\ref{join3}), the
requirement that $g_{\chi\chi}$ be $C^1$.  If
$g_{\chi\chi}$ is not $C^1$ at $\mathcal J$, then the
curvature will be a $\delta$ function on $\mathcal J$, but
this $\delta$ function will correspond to a shell-crossing
singularity, a singularity that is generally agreed to be
unphysical.  (Notice also that requiring $g_{\chi\chi}$ be
$C^1$ across $\mathcal J$ actually requires that the radii
$Y$ of the constant $\chi$ spheres have one of its {\it
second} derivatives, $\dot{Y}'$, be continuous across
$\mathcal J$.)  So we drop  the junction condition
(\ref{join3}).  A little manipulation yields

\begin{eqnarray}
\eta_{\mathcal J}-2\tan\frac{\eta_{\mathcal
J}}{2}=\frac{f^3(t_f-t_0)}{m}\label{cjoin2}\\ t_{\mathcal
J} = t_0(\chi) + (\eta_{\mathcal J} -\sin\eta_{\mathcal
J}) \frac{m(\chi)}{f^3(\chi)}\label{cjoin3}\\ N(\chi) =
\frac{|Y'(\eta_{\mathcal J}, \chi)|}{(t_f - t_{\mathcal
J}) \sqrt{1-f^2}}\label{cjoin1}\\ Z(\chi) =
|\dot{Y}(\eta_{\mathcal J}, \chi)|\label{cjoin4}
\end{eqnarray}

We proceed as in the previous section, solving first for
$\eta_{\mathcal J}$ and then substituting this into the
other equations.  Note that in order to invert
equation (\ref{cjoin2}), solving for
$\eta_{\mathcal J}$, the fact that $t_0$ must be less than
$t_f$ (the final $t$-value of the universe) implies that
the LHS should be positive for all values of $\chi$.  But
then since the positive values of $u-2\tan(u/2)$ for $u
\in [0,\,2\pi]$ are all at least $2\pi$, we must therefore
have the RHS being at least $2\pi$ for all values of
$\chi$.  Since $m$, $f$, and $t_0$ are all defined on the
same compact interval, we may (and must) choose the
constant $t_f$ so large that the RHS is always at least
$2\pi$.  Notice that the differentiability of $N$ in
equation~\ref{cjoin1} will be guaranteed when $Y' =0$
and $f=1$ because the value of $(Y')/(1-f^2)$ as $\chi$
approaches such a point is well defined and positive,
and thus $(t_f-t_0)$ will be just the square root of this
positive value.  This means that the more standard type
of shell-crossing singularity ($Y'=0$ but $f\not= 1$, so
that $g_{\chi\chi} =0$) is assumed not to occur on
$\mathcal J$.  For this reason, we called the allowed
singularity a ``weak'' shell-crossing singularity.

Therefore, if we make the appropriate choice of $t_f$ as
described above, we may join an arbitrary Tolman-Bondi
closed dust metric to the metric (\ref{NZ}) provided we
allow for a possible shell-crossing singularity.  The only
additional restriction we must impose on the Tolman-Bondi
functions $m$, $f$, and $t_0$ is that they must be chosen
to make the universe ``sufficiently large'' as discussed
in Section \ref{secNZ}.

\section{Black Holes}
\setcounter{equation}{0}

One interesting consequence of the above constructions
is that they provide examples of spacetimes satisfying
the energy conditions which can contain black holes,
but do not contain event horizons.  In order for this
statement to make sense, however, we need a good
definition of a black hole in a closed universe, for in a
closed universe the black hole singularity is actually
just a component of the final singularity (cf.
\cite{wheel}).  We will discuss in detail three such
definitions, the first due to Hayward \cite{hayward},
the second due to Tipler \cite{tipler2}, and the third
due to Wheeler \cite{wheel}.

In the standard definition of a black hole (cf. Wald
(\cite{wald1}, p. 300, or \cite{mtw}, p. 924), the black
hole $B$ is the spacetime region $B \equiv M -
J^-(\mathcal I^+)$, where $M$ is the spacetime
manifold, $\mathcal I^+$ is ``scri plus'' --- future null
infinity, and $J^-(S)$ is the {\em causal past} of a set
$S$, which is to say that $J^-(S)$ is the set of all
spacetime points $p$ which can be reached by a
past-directed timelike or null curve from $S$ to $p$. 
(Discussions of global general relativity and the
definitions of concepts used in this discipline can be
found in Wald (\cite{wald1}, chapter 8; Misner, Thorne
and Wheeler \cite{mtw}, Chapter 34; or (\cite{hawk})
and (\cite{penrose}).)  This definition cannot be applied
in a closed universe, because $\mathcal I^+$ does not
exist in a closed universe with a final singularity. 
However, this standard definition of a black hole is
never used in practice.  When astrophysicists search for
black holes, they look for gravitational fields implying
the presence of trapped or marginally trapped
surfaces.  In asymptotically flat spacetimes, (1) all
trapped surfaces can be proven to be inside of a black
hole (in the standard definition), and (2) black holes are
expected to evolve rapidly to a Schwarzschild or Kerr
black hole, in which there are trapped surfaces
arbitrarily close to the boundary of the black hole
$\partial J^-( \mathcal I^+)$ --- the event horizon.  Now
trapped surfaces {\em can} be in closed universes.

\subsection{Trapped and Marginally Trapped Surfaces}

Thus, the fundamental concept in Hayward's and
Tipler's definitions of a black hole is that of a trapped
surface (cf. Hayward \cite{hayward}). Let $\mathcal S$
be a compact spacelike 2-surface embedded in our
spacetime manifold.  Let $P$ be a point of $\mathcal
S$.  There are precisely two null directions normal to
$\mathcal S$ at $P$.  Suppose furthermore that these
null directions can be expressed as two vector fields
$N_+$ and $N_-$ defined on all of $\mathcal S$.  We
can choose both of $N_+$ and $N_-$ to be
future-directed.  Now, allowing $\mathcal S$ to evolve
along $N_+$ and $N_-$, we can measure its area at
every instant, and logarithmically differentiate the
resulting function with respect to the evolution
parameter.  Call these quantities $\theta_+$ and
$\theta_-$, respectively.

{\definition $\mathcal S$ is called a {\em\bf (future)
trapped surface} if both $\theta_+<0$ and $\theta_-<0$. 
If one of these quantities is zero and the other is
negative, then $\mathcal S$ is called a {\em\bf
marginally trapped surface}.}\\

The intuition here is that light rays emitted from a
trapped surface will converge, no matter whether they
are sent ``outward'' or ``inward.''  This is certainly a
necessary property of a black hole, and it would be
sufficient if it weren't for the fact that the
cosmological singularity produces a wealth of trapped
surfaces as the universe collapses.  In a FRW closed
universe, for example, there is a trapped surface
passing through every spatial point in the collapsing
phase.  In order to distinguish these cosmological
trapped surfaces from non-cosmological ones --- black
hole type trapped surfaces --- we need additional
criteria.

\subsection{Hayward's Black Hole Definition}

Marginally trapped surfaces are important because in
some sense they are where the horizon of a prospective
black hole should be.  To distinguish trapped surfaces
arising from black holes from those arising from the
collapse of the universe, therefore, Hayward considers
what should be happening in a neighborhood of a
marginally trapped surface which arises because of a
black hole.  Without loss of generality let $\theta_+=0$,
$\theta_-<0$ along the marginally trapped surface
$\mathcal S$.  Then $N_+$ can sensibly be called
``outward,'' $N_-$ ``inward.''  Outward-directed light
rays run instantaneously parallel to the surface, and
inward-directed light rays converge.  In the case of
black hole-based marginally trapped surfaces, however,
we would like to say that outward light rays just
outside $\mathcal S$ diverge, while outward light rays
just inside $\mathcal S$ converge. This can be
accomplished mathematically by extending the
embedding of $\mathcal S$ to a ``double-null foliation''
(cf. Hayward \cite{hayward}) in the direction of $N_+$
and $N_-$, extending $\theta_+$ appropriately, and
computing the sign of $\mathcal L_-\theta_+$, where
$\mathcal L_-$ denotes the Lie derivative in the
direction of $N_-$.

{\definition A marginally trapped surface with
$\theta_+=0$ (resp. $\theta_-=0$) is called {\em\bf
inner} if $\mathcal L_-\theta_+$ (resp. $\mathcal
L_+\theta_-$) is positive, {\em\bf outer} if $\mathcal
L_-\theta_+$ (resp. $\mathcal L_+\theta_-$) is
negative, and degenerate otherwise.}\\

As makes sense, inner marginally trapped surfaces
correspond to cosmological collapse, and outer
marginally trapped surfaces correspond to
non-cosmo\-logical collapse, i.e., to the marginally
trapped surfaces we would expect to find inside black
holes.  

As mentioned above, in asymptotically flat spacetimes,
all the trapped surfaces are inside the black hole, and
furthermore, all future directed causal (timelike or
null) curves from any trapped surface $\mathcal T_i$
can also be shown to be inside the black hole.  Thus if
$B$ is the black hole region, we must have $J^+(\cup_i
T_i) \subset B$ in order to capture the astrophysically
defining feature of a black hole in a black hole
definition applicable to a closed universe.  Also, in
asymptotically flat spacetimes, any spacetime point
$p$ whose causal future eventually enters the causal
future of a trapped surface can be proven to be inside
a black hole.  This means that we should also include
in the black hole $B$ all points $p$ such that
$J^+(J^+(p)\cap J^ -(\cup \mathcal T_i) )\subset J^-(\cup
\mathcal T_i)$.

This gives

{\definition {\bf (Hayward)} a {\em\bf black hole} is
the set of all spacetime points $p$ such that
$J^+(J^+(p)\cap J^-(\cup \mathcal T_i) )\subset J^-(\cup
\mathcal T_i)$. where $\cup \mathcal T_i$ is the union
of all outer marginally trapped surfaces}

\subsection{Tipler's Black Hole Definition}

Tipler's criterion (\cite{tipler2}) is related to
Hayward's, but perhaps is a bit simpler (see Hayward
\cite{hayward} for a short discussion of how these
criteria relate).

Instead of using a double-null foliation to test whether
a given marginally trapped surface corresponds to the
cosmological collapse or to a local black hole, Tipler
instead supposes that the marginally trapped surface in
question is contained in the boundary of a spacelike
hypersurface-with-boundary $\mathcal T$ whose interior
$\mathcal T-\partial \mathcal T$ is foliated by trapped
surfaces.  He then (assuming without loss of generality
that $\theta_+=0$, $\theta_-<0$) makes the following

{\definition If the family of null vectors $N_-$ (which
are all on $\partial \mathcal T$) point in the direction
of $\mathcal T$, then all trapped surfaces which can be
obtained from trapped surfaces in $\mathcal T$ by an
acausal homotopy foliated by trapped surfaces will be
called {\em\bf non-cosmological}.}

In particular, any trapped surface in $\mathcal T$ is
non-cosmological in this case.  Thus black hole type
trapped surfaces would be non-cosmological trapped
surfaces, and we have

{\definition{\bf (Tipler)} a {\em\bf black hole} is the
set of all spacetime points $p$ such that $J^+(J^+(p)\cap
J^-(\cup \mathcal T_i) )\subset J^-(\cup \mathcal T_i)$,
where $\cup \mathcal T_i$ is the union of all
non-cosmological trapped surfaces}

\subsection{Hayward-Tipler Black Holes in Tolman-Bondi
Closed Universes}

Specializing to the Tolman Dust case, we already have a
convenient foliation by 2-spheres, and we will use this
foliation to assist in evaluating the two definitions
outlined in the previous section.

First of all, note that the normal bundle to this
foliation is spanned by the vector fields
$\partial/\partial t$ and $\partial/\partial\chi$. 
Therefore we may set $$N_\pm=\frac{\partial}{\partial
t}\pm\frac{\sqrt{1-
f^2}}{Y'}\frac{\partial}{\partial\chi},$$  since
$\partial/\partial t$ is future-directed and
$g(\partial/\partial t,N_\pm)=1$.  Since the area of the
2-sphere at $(t,\chi)$ is $4\pi Y^2$, we can then easily
compute $$\theta_\pm=N_\pm(\log (4\pi
Y^2))=\frac{2f^3}{(1-
\cos\eta)m}\Big(\cot\frac{\eta}{2}\pm\frac{\sqrt{1
-f^2}}{f}\Big).$$ Note that in order for the 2-sphere at
$(t,\chi)$ to be marginally trapped, we are forced to have
$\theta_+=0$ and $\theta_-<0$ (since $\theta_+>\theta_-$).

Let us consider first Tipler's definition.  Choose
$\mathcal T$ to be a constant-$t$ hypersurface.  An
example of a vector pointing in the direction of $\mathcal
T$ is $\nu=-(\theta_+)'(\partial/\partial\chi)$ (here
primes are as in the join conditions above), since
$\theta_+$ will decrease to become negative on $\mathcal
T$, where the 2-spheres will be trapped surfaces.  Thus a
sufficient condition for cosmological trapped surfaces in
a neighborhood of a marginally trapped surface at
$(t,\chi)$ is
$$0<g(\nu,N_-)=(\theta_+)'\frac{Y'}{\sqrt{1-f^2}}.$$ 
Expanding this a bit, letting $Q=m/f^3$, ignoring positive
multipliers and using the fact that $\theta_+=0$, we
obtain the condition
$$\frac{Y'}{\sqrt{1-f^2}}\Big[\frac{t'_0+(\eta -
\sin\eta)Q'}{4f^4Q}- \frac{f'}{f^2\sqrt{1-f^2}}\Big]>0.$$

We now consider how Hayward's definition applies.  We are
taking the derivative in the $N_-$ direction of
$\theta_+$, and determining its sign.  To that end, we
will have a Hayward black-hole-type marginally trapped
surface if $$0 > \mathcal L_-\theta_+ =
\frac{1}{Y'}\Big\{\frac{1}{2f^3Q}\Big(\frac{f'}{f} -
\frac{Q'}{Q}\Big) - \frac{\sqrt{1 - f^2}}{2f^6Q^2}[t'_0 +
(\eta-\sin\eta)Q']\Big\}.$$

\subsection{Black Holes in a Joined Universe}

Now let us suppose that we are looking for a black hole
in a pressureless dust universe which can be joined to the
$N$-$Z$ universe defined above.  We first consider the
case of the differentiable join. Leaving $m$ and $f$ free
as in our derivation of the join conditions, we compute
Tipler's criterion to be 
$$\frac{Y'}{\sqrt{1-f^2}}\Big[\frac{(T+\eta -
\sin\eta)Q'}{4f^4Q}- \frac{f'}{f^2\sqrt{1-f^2}}\Big]>0,$$ 
and Hayward's to be $$\frac{1}{2Y'f^3Q}\Big\{\frac{f'}{f}
- \frac{Q'}{Q} - \frac{Q'(T + \eta-\sin\eta)\sqrt{1 -
f^2}}{Qf^3}\Big\}<0,$$  where $T(\chi)$ is defined
$$T=\frac{3\sin\eta_{\mathcal J}}{2+\cos\eta_{\mathcal
J}}-\eta_{\mathcal J}.$$

It is clear that we can choose $m,f,t_f$ in the $C^1$
join in such a way that the resulting joined universe
contains black holes in either the Hayward or Tipler
sense, and satisfies the energy conditions.  

\subsection{Wheeler's Black Hole Definition}

Following Wheeler and Qadir \cite{wheel}, we will
consider black holes in the spherically symmetric dust
universe to be re\-gions in which the universe is
collapsing much faster than elsewhere. An intuitive
measure of the rate of collapse of the universe in any
region is a measurement of the elapsed time between the
initial and final singularities.

We can use physical models to approximate the elapsed
time between singularities in black hole regions. For
simplicity, we will consider a one solar mass black hole.
Misner, Thorne, and Wheeler \cite{mtw} give the time
for a particle to fall from radius $R_i$ to the singularity
in a standard Schwarzschild black hole as
$\pi(R_i^3/8GM)^{1/2}$. Taking $R_i$ to be the
Schwarzschild event horizon $R_i = 2GM/c^2$, we have
that the elapsed time from the penetration of the event
horizon to the final singularity is approximately
$t_{horizon} = 5 \times 10^{-6}{\rm sec}(M/M_\odot)$. 
A dust cloud generating such a black hole would have a
total lifetime of twice this, giving us the elapsed time
from initial to final singularity as

\begin{equation}t_{total} = 10^{-5}{\rm sec}
\Big(\frac{M}{M_\odot}\Big)\label{BHlife}
\end{equation}

\noindent
in the vicinity of a black hole.

Misner, Thorne and Wheeler \cite{mtw} cite the {\em
illustrative} time that might be expected to elapse from
the beginning to the end of a typical closed FRW universe
(without black hole regions)as  $$t_{total} = 60 \times
10^9 \; \mathrm{years} = 1.8 \times 10^{17}\
\mathrm{sec}.$$ Combining this result with the above
calculation, we conclude that the elapsed time between
initial and final singularities in the vicinity of a 1
$M_\odot$ black hole will be on the order of $10^{22}$
times smaller than that in non-black hole regions of the
universe.  Since an upper bound to the mass of a black
hole in the current epoch of universe history is
believed to be $10^{10}M_\odot$, we would expect
that the elapsed time between the initial and final
singularities inside the largest black hole in existence
today would be on the order of $10^{12}$ times smaller
than that in non-black hole regions of the universe,
since by (\ref{BHlife}), a black hole lifetime scales
linearly with its mass.  

The elapsed time from the beginning to the end of the
universe along any timelike curves of constant $\chi$
can be obtained from (\ref{tdef}) by computing
$t(2\pi,\chi) - t(0,\chi)$. Thus the total time elapsed
from initial to final singularity along a timelike curve of
constant $\chi$ is simply $2\pi m(\chi)/f(\chi)^3$.
Therefore we can assert that if $\chi_a$ corresponds to
a black hole region of some hypersurface and $\chi_b$
corresponds to the cosmological region, we should
obtain that:  

\begin{equation} \label{intubh}
\frac{t_{total}(\chi_b)}{t_{total}(\chi_a)}=
\frac{m(\chi_b)/f(\chi_b)^3}{m(\chi_a)/f(\chi_a)^3} >
10^{13}.
\end{equation}

To apply this notion of black holes in the dust universe
to our joined metric, we consider any 2-sphere with
coordinate radius $\chi_a$ to be inside a black hole if
(\ref{intubh}) is satisfied when $\chi_b$ is the
coordinate radius of a 2-sphere whose size evolves
like the 2-spheres of spherical symmetry in a FRW
universe.   It will be sufficient that there exist a pair of
2-spheres with coordinate radii $\chi_a$ and $\chi_b$
such that (\ref{intubh}) is satisfied.  Alternatively,
we could simply restrict attention to $1 M_\odot$
black holes and fix $\chi_b$.  An elementary calculation
shows that a solar mass black hole is a 2-sphere with
radius corresponding to the radial value $\chi
= 10^{-23}$.  A third way of picking the pair of
2-spheres is to require that the $\chi_a$ 2-sphere is
the ``largest" 2-sphere in the universe ``today''.  This
would mean that the $\chi_a$ 2-sphere is an extremal 
--- maximal --- 2-sphere embedded in the 3-sphere
corresponding to ``today''.  We have to require the
2-sphere to be extremal  since one can construct a
non-extremal 2-sphere of arbitrary size embedded in a
3-sphere.

Wheeler points out \cite{wheel} that the natural
meaning of ``today'' --- the choice of a spacelike
hypersurface though the Earth --- is the constant mean
curvature hypersurface through the Earth today.  Tipler
(\cite{tipler3}, p. 440) has shown that if the strong
energy condition holds, and if the universe began close
to homogeneity and isotropy, an Omega Point spacetime
can be uniquely foliated by constant mean curvature
hypersurfaces, so Wheeler's proposal does indeed define
a unique ``today'' over the entire universe (Tipler also
shows that a constant mean curvature hypersurface
probably coincides with the rest frame of the CBR at any
event, so Wheeler's ``today across the entire universe'' is
even easy to locate experimentally.)  Putting all of these
criteria together yields

{\definition{\bf (Wheeler)} a {\em\bf black hole} is the
set of all spacetime points $p$ such that $J^+(J^+(p)\cap
J^-(\mathcal S) )\subset J^-(\mathcal S)$, where
$\mathcal S$ is any 2-sphere with coordinate radius
$\chi_a$ for which (\ref{intubh}) holds when $\chi_b$
is the coordinate radius of a maximal 2-sphere in the
constant mean curvature hypersurface which includes
the 2-sphere with coordinate radius $\chi_a$.}

It is clear that since (\ref{intubh}) does not depend on
the function $t_0$, just on the functions $m, f$, it is
possible to construct even in the case of the $C^1$ join
a universe which is essentially a closed dust FRW
everywhere outside a small black hole by Wheeler's
definition, a black hole which is centered at the origin
of coordinates $\chi =0$.

\section{Quintessence and Recollapse}
\setcounter{equation}{0}

The ``no event horizon solution'' to the black hole
information problem requires that the universe
recollapse to a final singularity before black holes
have time to evaporate.  However, the best
observations \cite{Ostriker}, independently confirmed
by a number of groups, indicate that the universe is
currently accelerating.  Furthermore, the observed
structure is best explained (given a Hubble constant of
$65 \pm 5$ km/sec-Mpc and spatial flatness) via a
$\Lambda$CDM model \cite{Turner1}.  If this
acceleration were to continue --- as it would if it were
due to a positive cosmological constant --- then the
universe would expand forever, and our proposed
solution to the BH unitarity problem would be incorrect:
unitarity would be violated.

However, Barrow (\cite{Barrow1}, see also \cite{Burd1})
was the first to point out that an accelerating universe
today need not preclude a recollapse in the far future of
a closed universe.  Since the acceleration of the scale
factor $R$ is given by equation (\ref{2ndfriedeq}),
namely $\ddot R = -(4\pi G/3)R(\rho +3p) = -(4\pi
G/3)R(1 + 3w)\rho$,  acceleration today implies that
$w <- 1/3$ today (the data give $w< -5/9$ today at the
95\% confidence level \cite{Turner2}), but if eventually
$w > -1/3$ for all time greater than some far future
value $t_{future}$, then the recollapse theorems of
Barrow, Galloway and Tipler \cite{tipler} will apply, and
recollapse will occur.

Thus unitarity implies that the observed acceleration
``today'' (meaning most of past {\it proper}
time) cannot be due to a positive cosmological
constant, but must instead be due to quintessence. 
This is of course the general expectation of
cosmologists, since the only plausible non-zero values
of the cosmological constant are near the Planck
density of $(10^{19}\,{\rm GeV})^4$, or near the density
of the SM Higgs field at its minimum $\sim (200\,{\rm
GeV})^4$, whereas the observed density of the material
causing the acceleration is of the order of the closure
density, $(10^{-3}\,{\rm eV})^4$ (\cite{Carroll1},
\cite{Weinberg1}).

The ``standard model'' of quintessence (\cite{Turner2},
\cite{Carroll2}) is a scalar field $\phi$ with a very
shallow potential
$V(\phi)$ in the present epoch, resulting in scalar field
excitations of very small mass, $m_\phi \equiv
\sqrt{V''(\phi)/2} \leq H_0 \sim 10^{-33}\,{\rm eV}$. 
Since we know very little more than this about
$V(\phi)$, the potential could have a minimum around
which the field will oscillate in the far future.  In such
a case, in the far future the leading term in the
expansion of the potential about the minimum would be
${1\over2}m^2_\phi\phi^2$, yielding \cite{Turner3} an
oscillation frequency $\omega = m_\phi$ and
$w\rightarrow0$ in the far future where $m_\phi >>
H(t)$.  With such a potential for the quintessence,
recollapse would occur, since the curvature term in the
Friedmann equation decreases as $R^{-2}$ whereas the
quintessence term would eventually decrease like the
matter, $R^{-3}$.  

Or the potential could be an exponential with no
minimum.  These are the most popular current models
of quintessence, since such potentials are suggested
by supersymmetry.  There are many exponential
potentials which allow recollapse, as established by
Barrow \cite{Barrow1}.  For example, if the potential
dies off sufficiently fast with $\phi$, then in the far
future, the density will drop off as $R^{-6}$ as does a
massless scalar field, the universe will become matter
and then curvature dominated in the far future, and
recollapse will result.

In summary, there are many quintessence models
consistent with all current observations which allow
recollapse in the far future.  Thus the scenario of
horizon elimination proposed in this paper is
consistent with all current astronomical observations.

\section{Conclusion}\label{conc}
\setcounter{equation}{0}

The ``holographic principle'' (\cite{birm1},
\cite{easther1}, \cite{pkraus1}) claims that all physics
on a manifold ---especially quantum gravity --- can be
completely described by a theory defined only on the
boundary of that manifold.  This is a completely
reasonable principle in the case that the boundary of the
manifold is a Cauchy surface for the manifold, because
in this situation the data on the boundary uniquely
determines the manifold and the properties of all
physical fields defined on the manifold.  For a classical
black hole which forms by collapse in an asymptotically
flat spacetime and then settles down to a Schwarzschild
exterior in the far future, the black hole event horizon
is indeed a Cauchy surface for the interior.  More
generally, if the spacetime is globally hyperbolic, we
would expect the event horizon to still be a Cauchy
surface for the black hole interior.  If we include the
c-boundary points in anti deSitter space, then the
Cauchy horizons surrounding a region plus the points
on the c-boundary where the horizon generators
terminate form a Cauchy surface for the interior
spacetime region enclosed by the Cauchy horizons; once
again we would expect the holographic principle to be
valid.

But there are problems with the holographic principle
in the case of black holes which evaporate to
completion.  Since the entire spacetime is no longer
globally hyperbolic, it is not clear that the event
horizon is a Cauchy surface for the interior.  There are
problems with the c-boundary completion:  looked
at from inside a black hole, the c-boundary
inside a spherically symmetric black hole is a 2-sphere
(the TIPs define a 2-sphere), whereas looked at from
the future after the evaporation is complete, the
c-boundary is a single point (the TIFs define a point). 
That is, the causal completion does not define the
boundary of the interior manifold uniquely.  Even if the
event horizon were actually a Cauchy surface for the
black hole interior, the information can never leave the
horizon to the exterior spacetime, since the event
horizon generators must terminate at the singularity
which ends the black hole evaporation.

This problem is obviated in an Omega Point spacetime. 
The null generators of the black hole apparent horizon
will actually be a Cauchy horizon for the entire
spacetime, for it can be shown that $\partial I^+(p)$ is
a Cauchy surface for the entire spacetime for any point
$p$ in the spacetime (Lemma 1 in\cite{tipler3}, p. 436). 
Thus the holographic principle is true for all manifolds
which are future sets (sets for which $I^+(S) \subset
S$).  In particular, for all points $p_i$ we wish to include
in ``black holes'' (by any of the defintions given above),
the boundary $\partial I^+(\cup p_i)$ will be a Cauchy
surface for the spacetime, and so the holographic
principle will hold for the surfaces of black holes in
Omega Point spacetimes.

Another area of general relativity that is naturally
complemented by the no-event-horizons resolution of
the black hole information problem is the computation
of gravitational radiation from colliding black holes. 
Matzner {\it et al} \cite{matzner2} have noted that the
computer simulation of a black hole collision is much
simpler if characteristic evolution is used in the black
hole exterior, because in asymptotically flat spacetimes,
the characteristic formulation can be compactified.  In
an Omega Point spacetime, the characteristic
formulation is {\it automatically} compactified: the null
boundary $\partial I^+(p)$ of any point $p$ in an Omega
Point spacetime has been shown by Tipler to be a
compact Cauchy surface (\cite{tipler3}, p. 436), as we
pointed out above.  We conjecture that the calculation
would be even easier done in an Omega Point
background space , such as the spherically symmetric
Omega Point spacetimes exhibited in Sections 2 and 4. 
In an Omega Point spacetime, it is not necessary to add
the c-boundary points to compactify characteristic null
surfaces like $\partial I^+(p)$.  This is important,
because as York has recently emphasized, in general
coordinate systems, the initial value problem cannot be
well-posed in general relativity.  However, Tipler has
shown (\cite{tipler3}, p. 440) that Omega Point
spacetimes which satisfy the strong energy condition
and begin in a ``crushing'' singularity" (all FRW
singularities are of this type, as are all ``stable''
singularities) possess a unique foliation by constant
mean curvature hypersurfaces, and York has shown that
the initial data problem{\it is} well posed on such a
hypersurface.  (If the universe is currently accelerating,
the strong energy condition will not hold everywhere,
but nevertheless a constant mean curvature foliation
will still exist, (\cite{tipler3}, p. 439).  However, the
foliation may only be unique in the very early universe
and in the very late universe where the strong energy
condition will hold.)  As we have emphasized
repeatedly, we define black holes operationally in terms
of trapped surfaces, just as is done by the groups trying
to compute the amount of radiation emitted from
colliding black holes.  Locally, their calculations of the
black hole surfaces would be the same in asymptotically
flat space and in an Omega Point spacetime.  No
quantum effects would effect the location or the size or
the existence of trapped surfaces evolved in the black
hole collision calculations.

Finally, we point out that many of the well-known
difficulties associated with doing quantum field theory
in curved spacetimes disappear in Omega Point
spacetimes.  As we mentioned in Section 1, Omega Point
spacetimes necessarily are foliated by compact Cauchy
surfaces, and in spacetimes with compact Cauchy
surfaces --- i.e., in closed universes --- there exists a
natural unitary equivalence class of quantum field
theory constructions, specifically, those constructed
from all the Hadamard vacuum states (\cite{wald2},
p.96).  (Roughly speaking, a ``Hadamard state'' is one in
which the short distance singularity structure of the two
point function in curved space is the same as it is in
Minkowski space \cite{wald2}, pp. 92--95).  In
spacetimes without compact Cauchy surface, there are
no unitarily equivalent representations of the quantum
field algebra, and it was this fact which led many
relativists to give up the postulate of unitarity.  In an
Omega Point spacetime, it is not necessary to give up
unitarity.

It is not even necessary to give up the notion of
``particle'' or ``vacuum state'' in a curved Omega Point
spacetime, as many relativists have previously believed
(e.g. \cite{wald2}, p. 59 and p. 96).  The method of
Hamiltonian diagonalization (\cite{wald2}, p. 65) will
define a unique vacuum and Fock space with respect to
any given Cauchy surface, and we pointed out above
that a unique foliation of the spacetime by constant
mean curvature exists in a physically realistic Omega
Point spacetime (unique except {\it possibly} in the
periods where the universe is accelerating).  These
constant mean curvature hypersurfaces are the natural
``rest frame'' of the universe, and are the natural
corresponding frames to the global Lorentz frames in
Minkowski space.  In FRW universes, the constant mean
curvature hypersurfaces are the ``rest frames'' of the
CBR --- observers on worldlines normal to these
hypersurfaces would measure isotropic CBR
temperature.  With respect to such a unique global
foliation, the notion of ``particle'' and ''vacuum state'' is
defined and is unique in curved spacetime.

And such notions must be defined if quantum field
theory in curved spacetime is to be a legitimate low
energy limit of the (still unknown) quantum theory of
gravity.  Weinberg (\cite{weinberg2}, p. 2) has pointed
out that quantum field theories are regarded today as
``mere effective field theories,'' just low energy
approximations to a more fundamental theory. 
Quantum field theories are not themselves fundamental,
but we use them only because any relativistic quantum
theory will closely approximate a quantum field theory
when applied to particles at a low enough energy.  If
this is true, then quantum particles are more
fundamental than quantum fields, and thus a
semi-classical theory like quantum field theory in a
classical curved space background must contain a
natural definition of the more fundamental entity, the
particle.

In short, assuming the universe to end in a c-boundary
which is a single point --- assuming the universe to be
an Omega Point spacetime --- solves the black hole
information problem, allows the standard concepts of
relativistic quantum mechanics to be carried over into
curved spacetimes, simplifies the characteristic
initial value problem, and is consistent with all
astronomical observations.  The actual universe may
indeed be an Omega Point spacetime.


\begin{thebibliography}{99}

\bibitem{hawk} S. W. Hawking and G. F. R. Ellis,
{\it The Large-Scale Structure of Spacetime}
Cambridge University Press, Cambridge, England,
(1973). 

\bibitem{exact} D. Kramer, H. Stephani, E. Herlt,
M. MacCallum, and E. Schmutzer ed. {\it Exact Solutions
of Einstein's Field Equations,} Cambridge University
Press, Cambridge, England. 

\bibitem{tipler} J. D. Barrow, G. J. Galloway, and F. J.
Tipler, ``The Closed Universe Recollapse Conjecture''
Mon. Not. Roy. astr. Soc., 223, 835--844 (1986).

\bibitem{mtw} C. W. Misner, K. S. Thorne, and J. A.
Wheeler, {\it Gravitation}, Freeman, San Fransicso,
(1973).

\bibitem{wheel} J. A. Wheeler and A. Qadir, ``York's
Cosmic Time Versus Proper Time as Relevant to
Changes in the Dimensionless `Constants', K-Meson
Decay, and the Unity of Black Hole and Big Crunch'', in
{\it From SU(3) to Gravity}, eds. E. Gotsman and G.
Tauber, pp. 383--394, Cambridge University Press,
Cambridge (1985)

\bibitem{penrose} R. Penrose, {\it Techniques of
Differential Topology in Relativity,} Society for
Industrial and Applied Mathematics, Philadelphia, PA,
USA (1972) 

\bibitem{hayward} S. A. Hayward, ``General Laws
of Black Hole Dynamics'', Phys. Rev. {\bf D 49},
6467--6474 (1994).

\bibitem{tipler2} F. J. Tipler, ``Black Holes in
Closed Universes'', Nature, {\bf 270},  500--501 (1977).

\bibitem{sachs} R. Budic and R.K. Sachs, ``Deterministic
Spacetimes,'' Gen. Rel. Grav. {\bf 7}, 21--29 (1976). 

\bibitem{wald1} R. M. Wald, {\it General Relativity,}
University of Chicago Press, Chicago (1984).

\bibitem{wald2} R. M. Wald, {\it Quantum Field
Theory in Curved Spacetimes and Black Hole
Thermodynamics}, University of Chicago Press, Chicago
(1994). 

\bibitem{tipler3} Frank J. Tipler {\it The Physics of
Immortality}, Doubleday, New York, 1994.

\bibitem{Beken1} Jacob D. Bekenstein, ``Universal
Upper Bound on the Entropy-To-Energy Ratio for
Bounded Systems,'' Phys. Rev. {\bf D23} (1981),
287--298; Jacob D. Bekenstein and M. Schiller, ``Proof of
the Quantum Bound on Specific Entropy for Free Fields,''
Phys. Rev. {\bf D39} (1989), 1109--1115.  

\bibitem{Beken2} Jacob D. Bekenstein,
``Non-Archimedean Character of Quantum Buoyancy and
the Generalized Second Law of Thermodynamics,' 'Phys.
Rev. {\bf D60} (1999), 124010,
gr-qc/9906058, shows the claim in \cite{wald2}, that the
Bekenstein Bound is not an essential assumption in any
derivation of the generalized Second Law and hence
need not be true, is incorrect.

\bibitem{Beken3} Jacob D. Bekenstein, ``How Fast
Does Information Leak Out From a Black Hole?'' Phys.
Rev. Lett. {\bf 70} (1993), 3680--3683.

\bibitem{Susskind1} L. Susskind, ``The World
as a Hologram,'' J. Math. Phys. {\bf 36} (1995),
6377--6396

\bibitem{Susskind2} L. Susskind {\it et al},
``Information Loss and Anomalous Scattering,'' Phys.
Rev. {\bf D46}, 3435--3443; ``End Point of Hawking
Radiation,''{\bf D46}, 3444--3449 (1992); ``TASI
Lectures on the Holographic Principle,''
hep-th/0002044 ``Puzzles and Paradoxes about
Holography,'' hep-th/9902182.

\bibitem{Hawking1} S.W. Hawking, ``Breakdown of
Predictability in Gravitational Collapse,'' Phys. Rev. {\bf
D14}, 2460--2473 (1976); ``The Unpredictability of
Quantum Gravity,'' Comm. Math. Phys. {\bf 87} (1982),
395--415; ``Loss of Information in Black Holes,'' in {\it
The Geometric Universe}, Oxford University Press,
Oxford, (1996).

\bibitem{tHooft1} G. 't Hooft, ``The Black Hole
Interpretation of String Theory'', Nuclear Physics {\bf
B335} (1990), 138--154.; see also
gr-qc/9509050 and gr-qc/9310006.

\bibitem{Barrow1}  John D. Barrow,  ``Cosmic
No-Hair Theorems and Inflation'' Phys. Lett. {\bf B187}
(1987), 12--16. See also astro-ph/9506049.

\bibitem{Burd1}  Adrian B. Burd, and John D. Barrow,
``Inflationary Models With Exponential Potentials,''
Nucl. Phys. {\bf B308} (1988), 929--945.

\bibitem{Turner1}  Michael S. Turner and M. White,
``CDM Models with a Smooth Component'' Phys. Rev. {\bf
D56} (1997), R4439--4443. See also astro-ph/9703161.

\bibitem{Turner2}  Michael S. Turner,  ``Why
Cosmologists Believe the Universe is Accelerating,''
astro-ph/9904049.

\bibitem{Carroll1} Sean M. Carroll, William H. Press, and
Edwin L. Turner, ``The Cosmological Constant,'' Ann.
Rev. Astron. Astro. {\bf 30} (1992), 499--542.

\bibitem{Weinberg1}  Steven Weinberg,  The
Cosmological Constant Problem,'' Rev. Mod. Phys. {\bf
61} (1989), 1--23. See also astro-ph/9610044 and
astro-ph/0002387.

\bibitem{Ostriker}  Neta A. Bahcall, Jeremiah P. Ostriker,
Saul Perlmutter, and Paul J. Steinhardt, ``The Cosmic
Triangle: Revealing the State of the Universe,'' Science
{\bf 284} (1999), 1481--1488. See also
astro-ph/9804065.

\bibitem{Vilenkin}  A. Vilenkin, ``String--Dominated
Universe,'' Phys. Rev. Lett.  {\bf 53} (1984),
1016--1018.

\bibitem{Turner3}  Michael S. Turner ``Coherent Scalar
Field Oscillations in an Expanding Universe,'' Phys. Rev.
{\bf D28} (1983), 1243--1247.

\bibitem{Susskind3}  Thomas Banks, Leonard Susskind,
and Michael E. Peskin ``Difficulties for the Evolution of
Pure States into Mixed States,'' Nuclear Physics {\bf
B244} (1984), 125--134.

\bibitem{Carroll2}  Sean M. Carroll, ``Quintessence and
the Rest of the World: Suppressing Long Range
Interactions,'' Phys. Rev. Lett. {\bf 81} (1998),
3067--3070, astro-ph/9806099.

\bibitem{tipler4}  John D. Barrow and Frank
J. Tipler 1986 {\it The Anthropic Cosmological
Principle}, Oxford University Press, Oxford, 1986.

\bibitem{matzner1} Richard A. Matzner {\it et al,}
``Geometry of a Black Hole Collision,'' Science {\bf 270}
(1995), 941--947.  See especially the ``horizon
generators'' of a pair of colliding black holes, which
appeared on the cover of the issue of Science containing
this article (10 November 1995).  The null geodesics so
pictured will still exist, and will be astrophysically
indistinguishable from the null geodesics pictured here,
but they will not be the generators of a event horzion. 
See also gr-qc/0002076.

\bibitem{matzner2} Richard A. Matzner {\it et al,}
``Cauchy Characteristic Matching: A New Approach to
Radiation Boundary Conditions,'' Phys. Rev. Lett. {\bf
76} (1996), 4303--4306, gr-qc/9807047.

\bibitem{price1} Richard Price {\it et al,} ``Inspiraling
Black Holes: The Close Limit,'' Phys. Rev. Lett. {\bf 83}
(1999), 3581--3584, gr-qc/9905081.

\bibitem{birm1} D. Birmingham, ``Geometrical
Finiteness, Holography, and the BTZ Black Hole,'' Phys.
Rev. Lett. {\bf 82} (1999), 4164--4167,
hep-th/9812206.

\bibitem{easther1} R. Easther and D. Lowe,
``Holography, Cosmology, and the Second Law of
Thermodynamcs,'' Phys. Rev. Lett. {\bf 82} (1999),
4967--4970, hep-th/9902088.

\bibitem{pkraus1} P. Kraus {\it et al} ``Spacetime and
the Holographic Renormalization Group,'' Phys. Rev. Lett.
{\bf 83} (1999), 3605--3608, hep-th/9903190.

\bibitem{weinberg2} Steven Weinberg, {\it The
Quantum Theory of Fields, Volume 1}, Cambridge
University Press, Cambridge, 1995.


\end{thebibliography}
\end{document}